\documentclass[pdflatex,sn-basic]{sn-jnl}% Math and Physical Sciences Numbered Reference Style

\usepackage{setspace}
\usepackage{graphicx}%
\usepackage{multirow}%
\usepackage{amsmath,amssymb,amsfonts}%
\usepackage{amsthm}%
\usepackage{mathrsfs}%
\usepackage[title]{appendix}%
\usepackage{xcolor}%
\usepackage{textcomp}%
\usepackage{manyfoot}%
\usepackage{booktabs}%
\usepackage{algorithm}%
\usepackage{algorithmicx}%
\usepackage{algpseudocode}%
\usepackage{listings}%
\usepackage{mathtools}
\usepackage{array}
\usepackage{xcolor}
\usepackage{soul}      % robust strikeout
\usepackage{natbib}

\soulregister\citep7   % allow citations inside strikeout

\newif\ifrevisions
\revisionsfalse

\definecolor{reviewercolor}{RGB}{34,139,34}   % forest green
\definecolor{delcolor}{RGB}{180,60,40}   % soft brick red
\definecolor{finalcolor}{RGB}{0,0,0}

\newcommand{\del}[1]{%
  \ifrevisions
    {\textcolor{delcolor}{\st{#1}}}%
  \else
  \fi
}

\makeatletter 

\DeclareRobustCommand{\rev}[1]{%
    \ifx\protect\@typeset@protect  
        \ifrevisions 
            \textcolor{reviewercolor}{#1}% 
        \else 
            #1% 
        \fi 
    \else 
        #1% 
    \fi 
} 

\makeatother

\begin{document}

\title[Article Title]{Quantifying the effect of phenotype on clustering behaviour in melanoma: from monoculture to co-culture}

\author*[1,2]{\fnm{Nathan J.} \sur{Schofield}}\email{nathan.schofield@dtc.ox.ac.uk}

\author[2]{\fnm{Richard M.} \sur{White}}\email{richard.white@ludwig.ox.ac.uk}

\author[1]{\fnm{Ruth E.} \sur{Baker}}\email{ruth.baker@maths.ox.ac.uk}
\author[1,2]{\fnm{Helen M.} \sur{Byrne}}\email{helen.byrne@maths.ox.ac.uk}

\affil[1]{\orgdiv{Mathematical Institute}, 
    \orgname{University of Oxford}, 
    \orgaddress{
        \street{Andrew Wiles Building, Woodstock Road}, 
        \city{Oxford}, 
        \postcode{OX2 6GG}, 
        \state{England}, 
        \country{United Kingdom}}}
\affil[2]{\orgdiv{Ludwig Institute for Cancer Research}, 
    \orgname{University of Oxford}, 
    \orgaddress{
        \street{Old Road Campus Research Building, Roosevelt Drive}, 
        \city{Oxford}, 
        \postcode{OX3 7DQ}, 
        \state{England}, 
        \country{United Kingdom}}}

\abstract{Melanoma is an aggressive form of skin cancer. Survival rates are excellent if it is detected early but fall markedly if it metastasises. A key step in early tumour progression is the formation of cell clusters, which can promote metastasis. However, the mechanisms driving cell clustering, and the role of phenotypic heterogeneity in the dynamics of these clusters, remain poorly understood.
In this work, we propose a system of ordinary differential equations that models cluster formation dynamics within a coagulation–fragmentation–proliferation framework. 
Using Bayesian inference, we fit this model to \textit{in vitro} time-lapse microscopy data from \rev{both monoculture and co-culture experiments involving} two melanoma phenotypes---proliferative and invasive---to uncover the predominant mechanisms driving cluster formation and how these differ between phenotypes. Additionally, we provide preliminary insights into how clustering behaviour in co-cultures contrasts with that observed in monocultures.  
The model quantifies phenotypic differences in clustering dynamics: invasive cells in monoculture exhibit nearly threefold higher coagulation rates than proliferative cells, whereas proliferative cells display slightly higher proliferation rates. These differences align with known gene expression profiles. When applied to co-culture data, the model predicts hybrid coagulation behaviour of the clusters influenced by both proliferative and invasive cells but dominated by the invasive cells, and an elevated proliferation rate, suggesting a mutually beneficial effect of phenotypic heterogeneity on cell proliferation.}


\maketitle

\section{Introduction}
Melanoma is a form of skin cancer that arises from melanocytes, the pigment-producing cells responsible for skin colour \citep{cancerresearchuk_melanoma}. It is the fifth most common cancer in the UK \citep{nhs_melanoma}, with many cases being preventable as they are linked to ultraviolet (UV) radiation exposure. When detected early, melanoma has a favourable prognosis, with a 10-year survival rate exceeding 90\% in the UK \citep{pcds_melanoma}. However, prognosis worsens significantly once the cancer metastasises.
The formation of clusters of melanoma cells is known to promote metastasis \citep{aceto_circulating_2014,fidler_relationship_1973,gkountela_circulating_2019,glaves_correlation_1983,liotta_significance_1976,szczerba_neutrophils_2019}. Despite their clinical importance, the mechanisms driving cluster formation remain poorly understood, particularly in the context of tumour heterogeneity \citep{dagogo-jack_tumour_2018}. Understanding how different melanoma phenotypes contribute to clustering could reveal new strategies for disrupting metastasis and, in so doing, improve outcomes for patients with metastatic melanoma.

\rev{
From a mathematical perspective, tumour growth and invasion have been widely studied using continuum, reaction--diffusion, agent-based, and multiscale frameworks that capture processes including spatial invasion, cell migration, and metastatic spread
\citep{franssen_mathematical_2019,gatenby_reaction-diffusion_1996,hodgkinson_computational_2022}.
Within melanoma, mathematical and computational models have addressed intracellular signalling, treatment response, tumour--immune interactions, angiogenesis, and spatial growth and invasion
\citep{Albrecht2020ComputationalMelanoma,dePillis2013DendriticCellTherapy,Lai2017MelanomaCombinationTherapy,Wang2013MelanomaAngiogenesis}.
More recently, models have examined proliferative–invasive plasticity and phenotype-structured population dynamics that link the tumour-scale behaviour to cell states regulated by MITF, a transcription factor that governs melanocyte identity and phenotype switching \citep{Subhadarshini2023MelanomaPlasticity,Chambers2026MITFRheostat}.
However, comparatively little attention has been paid to the early formation of melanoma cell clusters and the evolution of their cluster-size distributions.
}

\rev{
Aggregation theory provides a natural framework for modelling the evolution of cluster-size distributions.
Classical Smoluchowski-type equations describe the formation of larger clusters through pairwise coagulation, with the coagulation kernel determining how merging rates depend on cluster size.
Coagulation--fragmentation extensions additionally account for cluster fragmentation and the resulting competition between assembly and disassembly
\citep{aktar_finite_2021,kuehn_smoluchowskis_2019,laurencot_weak_2015,suzuki_predictions_1969,wattis_introduction_2006}.
However, to our knowledge, these mechanistic frameworks have not been calibrated to time-resolved melanoma cluster-size distributions to distinguish the contributions of cluster merging, cell proliferation, and cluster fragmentation, or to compare the inferred dynamics across melanoma phenotypes and between monoculture and co-culture conditions.
}

\rev{
Here, we address this gap by developing a coagulation--fragmentation--proliferation model and calibrating it to time-resolved phenotype-specific data.}
By combining parameter inference with mechanistic model comparison, the framework enables us to identify the processes required to explain cluster formation and to quantify differences in the inferred clustering dynamics across phenotypic conditions within a unified, data-driven framework.

\del{To address this, }\rev{To implement this framework, we} analyse previously published \textit{in vitro} data \citep{campbell_cooperation_2021} showing cluster formation for two distinct melanoma phenotypes which we refer to as proliferative and invasive. These phenotypes are well-characterised and mutually exclusive at the single-cell level \citep{tirosh_dissecting_2016,tsoi_multi-stage_2018,widmer_systematic_2012,wouters_robust_2020}. The dataset includes monoculture experiments for each phenotype and a co-culture experiment where the two phenotypes interact. 

We develop a system of ordinary differential equations (ODEs) to describe clustering dynamics. The model consists of three key mechanisms: coagulation (clusters merging), fragmentation (clusters splitting), and proliferation (clusters growing in size through cell division).
The aim of this modelling work is to determine which mechanisms and mathematical representations of these mechanisms are required to reproduce the observed clustering dynamics, and how these requirements differ between phenotypes. To explore mechanistic hypotheses about phenotypic differences in clustering, we construct a family of model variants that differ in their functional representations of proliferation and fragmentation. We use parameter inference and model selection to fit these model variants to experimental data, which enables us to gain mechanistic insights into clustering behaviour and expose \rev{and quantify} differences in behaviour between the two phenotypes. We also evaluate the behaviour of heterogeneous (co-cultured) clusters, comparing them to monocultures and identifying where phenotypic mixing alters clustering dynamics.

The remainder of this paper is organised as follows. Section \ref{section:Methods} describes the development of the mathematical model, the analytical techniques employed, and the approach used to fit the model to experimental data. Section \ref{section:Results} presents the outcomes of the model analysis, beginning with a comparison of clustering dynamics in monocultures of proliferative and invasive cells, followed by an investigation of co-culture behaviour. Finally, in Section \ref{section:Discussion}, we discuss the results, their implications, and outline potential directions for future research.

\section{Methods \label{section:Methods}}

In this section, we introduce the ODE model \rev{used to represent the biological processes that give rise to the observed data.}\del{that we use to describe the data.} We also introduce the different model variants and describe the Bayesian framework we use for parameter estimation and model selection. We start by briefly describing the experimental data we use to fit the model.

\subsection{Experimental data \label{section:ExperimentalDataOverview}}

We analyse existing data \citep{campbell_cooperation_2021} from two melanoma cell phenotypes of the ZMEL1 line \citep{heilmann_quantitative_2015}. One phenotype exhibits higher proliferative capacity, while the other is more invasive. The dataset includes experiments on homogeneous populations of these proliferative and invasive phenotype cells, as well as co-culture experiments where proliferative and invasive cells are mixed in a $1{:}1$ ratio. In each case, cells were initially plated as single cells and, over time, formed clusters comprising multiple cells. Maximum intensity projection images were acquired for each experiment every $30$ minutes over a $72$-hour time period. 
The timestep used in the analysis is therefore defined as $\Delta t = 30\ \text{minutes}$, corresponding to the imaging interval. \rev{Since the experimental timestep is $\Delta t = 30$ minutes, all reported rate parameters are converted to per hour units (h$^{-1}$).}
These images were segmented to generate cluster size distributions throughout the experiment, which serve as the data for model fitting. Because maximum intensity projections reduce three‑dimensional cell clusters to two‑dimensional representations, we convert cluster areas to estimated volumes to recover a three‑dimensional measure of cluster size. Additional details on the experimental procedures and image analysis methods used to derive cluster size distributions are provided in the Supplementary Information.

\subsection{Mathematical model development \label{section:HomogeneousModel}}

We denote by $\psi_i(t)$ the number of clusters of size $i$ observed in an experiment at time $t$, where $i$ is an integer denoting the number of cells in a cluster. 
We describe the time evolution of the cluster size distribution using a coagulation--fragmentation--proliferation type framework \citep{krapivsky_kinetic_2010, smoluchowski_drei_1916}. 
We denote by $T_i^{\text{coag}}$ the net rate at which clusters of size $i$ form due to coagulation, $T_i^{\text{prolif},\alpha}$ the net rate at which they form due to cell proliferation under proliferation model $\alpha$, and $T_i^{\text{frag},\beta}$ the net rate at which they form due to fragmentation under fragmentation model $\beta$. By appealing to the principle of mass balance, we deduce that the time evolution of clusters of size $i$ can be written as

\begin{equation}
\frac{\text{d}\psi_i}{\text{d}t} = T_{i}^{\text{coag}} + T_{i}^{\text{prolif},\alpha} + T_{i}^{\text{frag},\beta}.
\label{eqn:Homo-master-eqn}
\end{equation}

\noindent \textbf{Coagulation.}
The coagulation term models the merging of two clusters to form a larger one. Motivated by the experimental data (see Supplementary Information Figure S2), we assume that clusters coagulate at a constant rate $b>0$ which is independent of their size.
The net coagulation rate $T_{i}^{\text{coag}}$ for $\psi_{i}$ comprises two terms: a loss term due to coagulation of clusters of size $i$ with other clusters and a source term due to the formation of clusters of size $i$ when smaller clusters of sizes $1 \leq k < i$ and $i-k$ coagulate. Under this assumption, the coagulation term, $T_{i}^{\text{coag}}$, can be written as:\\

\begin{equation}
T_{i}^{\text{coag}} = 
\begin{cases} 
-b\displaystyle\sum\limits_{j=1}^{\infty} \,\psi_1 \psi_j, & \text{if } i=1, \\
-b\displaystyle\sum\limits_{j=1}^{\infty} \,\psi_i \psi_j + \frac{b}{2}\displaystyle\sum\limits_{j=1}^{i-1} \psi_j \psi_{i-j}, & \text{if } i > 1.
\end{cases}
\label{Eqn:coag}
\end{equation}

\noindent \rev{This functional form assumes that all cluster pairs coagulate at the same rate, irrespective of the sizes of the two clusters, consistent with the experimental observation that the rate of coagulation does not depend strongly on cluster size (see Supplementary Information Figure S2).\\}

\noindent \textbf{Proliferation.}
Proliferation is the process by which a single cell divides into two cells. Here, proliferation is therefore the process by which a cluster increases in size by one cell because one of its constituent cells divides. We model proliferation as a growth process that depends on the size of the cluster, so clusters of size \(i\) transition to clusters of size \(i+1\) at a rate \(p \cdot f_i^\alpha\), where \(p > 0\) is a constant and \(f_i^\alpha\) is a functional form selected from a set of functions defined below. Under these assumptions, the proliferation term, $T_{i}^{\text{prolif},\alpha}$, indexed by $\alpha$ can be expressed as follows:

\begin{equation}
T_{i}^{\text{prolif},\alpha} = 
\begin{cases} 
-pf_1^\alpha \psi_1, & \text{if } i=1, \\
p\,(f_{i-1}^\alpha\psi_{i-1} - f_{i}^\alpha\psi_i), & \text{if } i > 1.
\end{cases}
\label{Eqn:prolif}
\end{equation}

\noindent \rev{This functional form assumes that cluster growth occurs via division of individual cells within the cluster, with the function $f_i^{\alpha}$ controlling how proliferation scales with cluster size.\\}

\noindent \textbf{Fragmentation.}
Fragmentation refers to the process by which a cluster breaks into multiple smaller clusters. Within the \textit{in vitro} data, however, we only observe a specific form of fragmentation in which a small cluster, typically of size one, detaches from the main cluster. We refer to this particular type of fragmentation as shedding. Accordingly, we model fragmentation exclusively as a shedding process, meaning a cluster of size $i > 1$ loses a cell and produces two clusters of sizes $i-1$ and $1$. We denote by $q \cdot f_i^\beta$ the rate at which clusters of size $i$ shed. As for proliferation, $q>0$ is a constant and $f_i^\beta$ is a functional form chosen from a set of functions defined below. Then we have the following expressions for the shedding term, $T_{i}^{\text{frag},\beta}$, indexed by $\beta$:

\begin{equation}
T_{i}^{\text{frag},\beta} = 
\begin{cases} 
q\,(f_2^\beta\psi_2\, + \sum\limits_{j=2}^{\infty} f_j^\beta\psi_j), & \text{if } i=1, \\
q\,(f_{i+1}^\beta\psi_{i+1} - f_i^\beta\psi_i), & \text{if } i > 1.
\end{cases}
\label{Eqn:shed}
\end{equation}

\noindent \rev{This expression reflects experimental observations that fragmentation events predominantly involve the detachment of single cells (see Supplementary Information Figure S2).\\}

\noindent Combining Equations (\ref{eqn:Homo-master-eqn})--(\ref{Eqn:shed}), we deduce that the following ODEs describe the cluster dynamics:

% \newpage

\begin{subequations}
    \begin{align}
        \begin{split}
        \frac{\text{d}\psi_1}{\text{d}t} =&-\underbrace{b\sum_{j=1}^{\infty} \,\psi_1 \psi_j}_{\text{Coagulation}}
        \qquad\qquad\qquad\quad\quad-\underbrace{pf_1^\alpha \psi_1\vphantom{\sum_{j+k=i}\frac{b}{2}\psi_j\psi_k}}_{\text{Proliferation}}\\
        &+\underbrace{q\,\left(f_2^\beta\psi_2\, + \sum\limits_{j=2}^{\infty} f_j^\beta\psi_j\right)\vphantom{\xi_{i+1}\psi_{i+1} - \xi_i\psi_i}}_{\text{Shedding (fragmentation)}};
        \end{split}\\
        \label{eqn:coag-shed-gen}
        \begin{split}
        \frac{\text{d}\psi_i}{\text{d}t} =&-\underbrace{b\sum_{j=1}^{\infty} \,\psi_i \psi_j + \frac{b}{2}\sum_{j=1}^{i-1}\psi_j\psi_{i-j}}_{\text{Coagulation}}
        \quad+\underbrace{p\,(f_{i-1}^\alpha\psi_{i-1} - f_{i}^\alpha\psi_i)\vphantom{\sum_{j+k=i}\frac{b}{2}\psi_j\psi_k}}_{\text{Proliferation}}\\
        &+\underbrace{q\,(f_{i+1}^\beta\psi_{i+1} - f_i^\beta\psi_i)\vphantom{\xi_2\psi_2 + \sum_{j=2}^{\infty} \xi_j\psi_j}}_{\text{Shedding (fragmentation)}} \quad\qquad\qquad\qquad\qquad\qquad\qquad\qquad \text{if } i > 1.
        \end{split}
    \end{align}
    \label{eqn:generalode}
\end{subequations}

Mirroring the experimental set-up, we assume that the system initially contains $N_{\rev{0}} > 0$ clusters of size one $(i=1)$. Thus we close Equation (\ref{eqn:generalode}) by imposing the following initial conditions:

\begin{equation}
    \psi_i(t = 0) = \begin{cases}
        N_{\rev{0}} \quad &i = 1,\\
        0 \quad &i > 1.
    \end{cases}
    \label{eqn:IC}
\end{equation}

Equations (\ref{eqn:generalode}) and (\ref{eqn:IC}) define an infinite system of ODEs describing how the cluster size distribution evolves over time. When simulating the model and fitting it to data, \del{we impose a maximum cluster size, $Y_{\max} = 100$, to truncate the system.}\rev{we impose a maximum cluster size, $i_{\max} = 100$, to truncate the system, where $i_{\max}$ denotes the largest cluster size included in the truncated system.} \\% This converts the infinite system into a finite set of ODEs.\\

\noindent \textbf{Model variants.}
To account for different mechanisms that drive proliferation and shedding, we introduce model variants by proposing different functional forms for $f_i^{\alpha}$ and $f_i^{\beta}$.
We denote a model variant as $\mathcal{M}_{\alpha,\beta},\,\, \alpha,\beta \in \{0,1,2,3\}$ where $\alpha$ denotes the functional form associated with proliferation and $\beta$ denotes the functional form associated with shedding. This gives $16$ model variants to compare.
The same set of functional forms is used for both proliferation and shedding, and these are defined as follows:

\begin{itemize}
    \item (None): $f_i^{0} := 0$. In this case, there is no proliferation/shedding.
    \item (Constant): $f_i^{1} := 1$. Here the rate of proliferation/shedding is independent of cluster size. 
    \item (Surface area): $f_i^{2} := i^{\frac{2}{3}}$. Here, we view the clusters as three-dimensional spheres and assume that the rate of proliferation/shedding is proportional to the surface area of the three-dimensional cluster.
    \item (Cell number): $f_i^{3} := i$. Here we assume that the rate of proliferation/shedding is proportional to the number of cells in the cluster.
\end{itemize}

\noindent To facilitate model analysis, we compute summary statistics, denoted by $S_h$, for the distributions of cluster sizes (see Supplementary Information Figure S1D). These statistics are defined as:
\begin{align*}
    S_0(t) &= \sum_{i = 1}^{\infty}\psi_i(t) = \text{Total number of clusters;}\\
    S_1(t) &= \frac{1}{S_0(t)}\sum_{i = 1}^{\infty}i\psi_i(t) = \text{Mean cluster size;}\\
    S_2(t) &= \frac{1}{S_0(t)}\sum_{i = 1}^{\infty}i^2\psi_i(t) = \text{Mean of the squared cluster sizes.}
\end{align*}

\noindent We also derive the ODE for the total number of cells over time by starting from the definition

\begin{equation*}
N_{\text{cells}}(t) = \sum_{i=1}^{\infty} i \cdot \psi_i(t).
\label{eqn:total_cells_def}
\end{equation*}

\noindent Differentiating both sides with respect to time gives

\begin{equation*}
\frac{\text{d}N_{\text{cells}}}{\text{d}t} = \sum_{i=1}^{\infty} i \cdot \frac{\text{d}\psi_i}{\text{d}t} = \sum_{i=1}^{\infty} i \, (T_{i}^{\text{coag}} + T_{i}^{\text{prolif},\alpha} + T_{i}^{\text{frag},\beta}).
\label{eqn:total_cells_derivative}
\end{equation*}

\noindent As both coagulation and shedding conserve the total number of cells in the system we have
\begin{equation}
\frac{\text{d}N_{\text{cells}}}{\text{d}t} = \sum_{i=1}^{\infty} p f_i^\alpha \psi_i,
\label{eqn:total_cells_final}
\end{equation}

\noindent closed with the initial condition
\begin{equation}
N_{\text{cells}}(0) = N_{\rev{0}}.
\label{eqn:total_cells_initial_condition}
\end{equation}

\subsection{Noise model \label{section:NoiseModel}}

We simulate the system of ODEs to obtain cluster size distributions $\psi_i(t_j)$ at discrete times $t_j = j\Delta t$, $\, j \in \{0,1,\dots, 144\}$.
From these, we calculate the summary statistics $S_h(t_j)$, $(h=0,1,2)$ which we view as ``observables" when fitting the model variants to the experimental data. These summary statistics define a lower-dimensional representation of the data, reducing the number of noise parameters needed to fit the ODE model to the data.
When fitting the summary statistics generated by the ODE model to the experimental data $(\boldsymbol{S}^\mathcal{D})$, which is noisy, we assume a log-normal noise model to preserve the strict positivity of the summary statistics, $S_0, S_1, S_2,$ so that

\begin{equation}
    S_h^{\mathcal{D}}(t_j) \;=\; S_h(t_j)\,\exp(\epsilon_h), 
    \quad \epsilon_h \sim \mathcal{N}(0,\sigma_h^2),
    \quad h \in \{0,1,2\}.
    \label{eqn:NoiseModel}
\end{equation}

\noindent \rev{Here, $\sigma_h$ ($h=0,1,2$) are noise parameters that are associated with the summary statistics $S_h$ and quantify the level of observational variability in each statistic.}

\subsection{Parameter estimation \label{section:ParameterEstimation}}

\rev{Before attempting to infer model parameters from experimental data, we first verify that the model is structurally identifiable \citep{chis_structural_2011,cobelli_parameter_1980}, since this ensures that parameter values can, in principle, be uniquely determined from ideal (noise-free) observations \citep{jensen_establishment_2015,raue_comparison_2014}.}\del{Before fitting the model variants to the data, we first establish that they are globally structurally identifiable \citep{chis_structural_2011,cobelli_parameter_1980}. A model is said to be globally structurally identifiable if each set of parameter values produces a unique distribution of model outputs \citep{jensen_establishment_2015,raue_comparison_2014}.} % \citep{audoly_global_1998,bellu_daisy_2007,chappell_structural_1992,ljung_global_1994,walter_identifiability_2013}. 
To assess global structural identifiability, we use the \texttt{GenSSI} package \citep{ligon_genssi_2018} and confirm that all parameters are structurally globally identifiable for each model variant. \rev{To verify that this result does not depend on the choice of truncation level, we repeat the analysis for $i_{\max} \in \{2, 3, 4, 5, 10, 20, 50, 100, 200\}$ and find that global structural identifiability is maintained throughout.} \del{This verification is repeated for a range of truncation values $i_{\max} \in \{2, 3, 4, 5, 10, 20, 50, 100, 200\}$ with global structural identifiability maintained across all cases.}

After establishing global structural identifiability, \rev{we proceed to parameter estimation using a Bayesian framework, which also allows us to assess practical identifiability. Practical identifiability concerns whether parameters can be estimated from finite, noisy data and is evaluated through the resulting posterior distributions \citep{falco_quantifying_2023}.}\del{we explore practical parameter identifiability, which concerns whether parameters can be estimated from finite, noisy data \citep{falco_quantifying_2023}.} Using established Bayesian methods \citep{gelman_bda3_2013,hines_determination_2014,simpson_practical_2020}, we compute posterior distributions, with broad or non-unimodal posteriors indicating \rev{parameters that are poorly constrained by the data \citep{siekmann_mcmc_2012}.} \del{poor identifiability \citep{siekmann_mcmc_2012}.}

To investigate \rev{practical identifiability,}\del{this,} we fit the model variants to the cluster size distribution data $(\boldsymbol{S}^{\mathcal{D}})$ extracted from experimental images, estimating parameters and uncertainty through the combined mechanistic process model, which represents cluster growth and shedding, and the noise model in Equation (\ref{eqn:NoiseModel}), which captures observational variability.
We associate with each model variant $\mathcal{M}_{\alpha,\beta}$ a parameter set $\boldsymbol{\theta} = \{b,p,q,N_{\rev{0}}, \sigma_0,\sigma_1,\sigma_2\}$ where $\boldsymbol{\theta}_M = \{b,p,q,N_{\rev{0}}\}$ are the deterministic model parameters and $\boldsymbol{\theta}_N = \{\sigma_0,\sigma_1,\sigma_2\}$ are noise parameters.
To infer the posterior distributions of these parameters, we utilise an MCMC approach with the PINTS package \citep{clerx_probabilistic_2019}, employing a prior that is uniform on the log-scale for each parameter.  
We use the Slice Sampling with Doubling method, as described in \citep{neal_slice_2003}. 
We use three chains and assess their convergence using the $\hat{R}$ diagnostic, as defined in \citep{gelman_bda3_2013}, with a convergence threshold of $1.05$.

\subsection{Model selection \label{section:ModelSelection}}

To compare alternative mechanistic model variants, we use the Akaike Information Criterion (AIC), which balances goodness of fit with model complexity by penalising models that include more parameters. This requires the computation of the log-likelihood function which, for a given parameter set $\boldsymbol{\theta}$, is defined as
$$
\ln \mathcal{L}(\boldsymbol{\theta} \mid \boldsymbol{S}^\mathcal{D}) = \sum_{i=1}^{n} \ln \big( P(S_i^{\mathcal{D}} \mid \boldsymbol{\theta }) \big),
$$
where $n$ denotes the total number of summary statistics across all time points, which is three times the number of time points analysed. The AIC is then calculated as
$$
\text{AIC} = 2k - 2\ln \mathcal{L}(\boldsymbol{\theta} \mid \boldsymbol{S}^\mathcal{D}),
$$
where $k$ denotes the number of estimated parameters, \rev{and the likelihood is evaluated at the maximum likelihood estimate of the parameters.}
We report relative information criterion values, that is $\Delta \mathrm{AIC}$ values, where $\Delta \mathrm{AIC}$ is the difference between a given model variants's AIC and that of the best fitting model variant (the model variant with the lowest AIC value).

\section{Results \label{section:Results}}

We apply the modelling framework developed in Section~\ref{section:Methods} to the cluster size distributions collected from the experimental data in order to determine the mechanisms driving melanoma cell clustering and quantify differences in these processes between proliferative and invasive phenotypes in monoculture and co-culture.

\subsection{Characterising the dynamics of clusters in monoculture \label{section:ResultsMonoculture}}

To compare the clustering behaviour of proliferative and invasive cells, we fit each of the $16$ model variants to the summary statistics derived from cluster size distributions over the course of the experiments. 
Because our focus is on the early stages of cluster formation, we restrict the analysis to timepoints where all observed clusters contain fewer than $100$ cells. Therefore, we truncate the infinite ODE system at a maximum cluster size of \rev{$i_{\max}=100$.} \rev{This choice reflects the range of cluster sizes observed in the experimental data.}
For illustrative purposes, we start with a case study of the parameter inference process for invasive cells using model variant $\mathcal{M}_{3,1}$.

\begin{figure}[H]
\centering
\includegraphics[width=\textwidth]{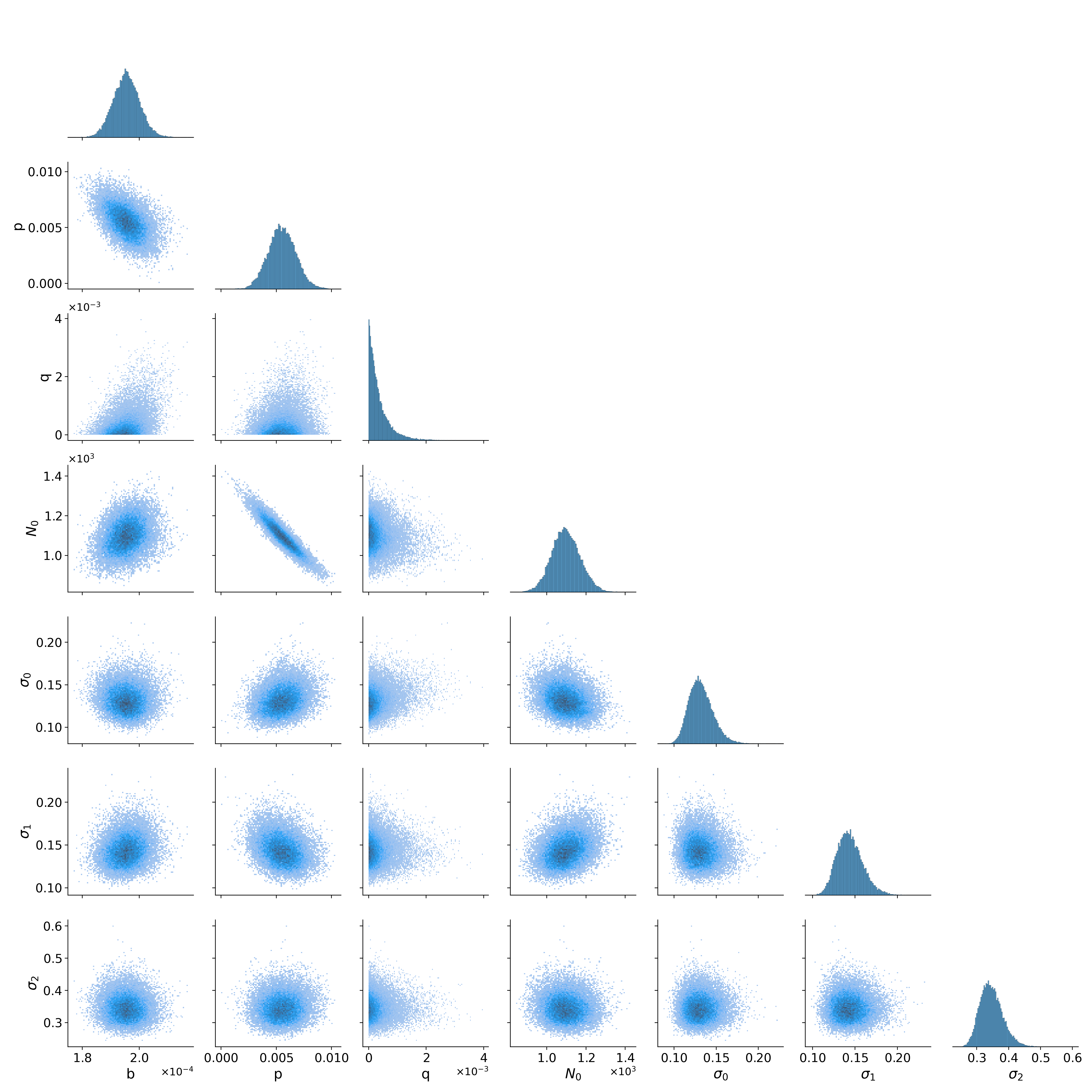}
\caption{Marginal and bivariate posterior distributions for the model parameters $\boldsymbol{\theta_M} = \{b,p,q,N_{\rev{0}}\}$ and noise parameters $\boldsymbol{\theta_N} = \{\sigma_0, \sigma_1, \sigma_2\}$  when MCMC is used to fit model variant $\mathcal{M}_{3,1}$ to summary statistics $S_0(t),S_1(t),S_2(t)$ extracted from the invasive cell data. \rev{Parameter values are shown with time measured in increments of $\Delta t = 30$ minutes. The coagulation rate $b$ has units of clusters$^{-1}\,\Delta t^{-1}$, the proliferation and shedding rates $p$ and $q$ have units of $\Delta t^{-1}$, and the initial number of clusters $N_0$ and noise parameters $\sigma_0, \sigma_1, \sigma_2$ are dimensionless.}}
\label{fig:INV_3_RAW_shed_pairwise_full}
\end{figure}

We recall that for $\mathcal{M}_{3,1}$ the proliferation rate is proportional to the number of cells in a cluster and the shedding rate is constant, independent of cluster size. Fitting this model variant to the invasive cell data yields the posterior distributions shown in Figure \ref{fig:INV_3_RAW_shed_pairwise_full}, which are unimodal and have $\hat{R}$ values less than $1.05$.

In Figure \ref{fig:INV_3_RAW_shed_pairwise_full} we also observe that there is a negative correlation between the proliferation rate, $p$, and the initial number of singleton cells, $N_{\rev{0}}$. We recall from Equations (\ref{eqn:total_cells_final}) and (\ref{eqn:total_cells_initial_condition}) that the total number of cells in the system only depends on two parameters: the proliferation rate, $p$, and the initial number of singleton cells, $N_{\rev{0}}$. This leads to the negative correlation observed in Figure \ref{fig:INV_3_RAW_shed_pairwise_full}, a behaviour that is conserved across cell type and choice of model variant (see Supplementary Information Figures S4-S9).
\rev{Additional, weaker correlations are observed between other parameters in some circumstances (see Supplementary Information Figures S4--S9).}
We also observe that the posterior distribution for the shedding parameter, $q$, is sharply peaked at zero, indicating that shedding by cluster of invasive cells is negligible and not required to explain the data.
No correlations are observed between the mechanistic parameters $\boldsymbol{\theta_M} = \{b,p,q,N_{\rev{0}}\}$ and the noise parameters $\boldsymbol{\theta_N} = \{\sigma_0,\sigma_1,\sigma_2\}$, nor among the noise parameters themselves. This indicates that the noise parameters are identifiable, meaning their individual effects on observational variability can be distinguished without confounding, which is important for reliable uncertainty quantification and interpretable parameter estimates \citep{raue_comparison_2014}.
We next performed a posterior predictive check to evaluate whether model variant $\mathcal{M}_{3,1}$ can reproduce the observed cluster size dynamics of the invasive cells \citep{berkhof_posterior_2000,gelman_bda3_2013,mcelreath_statistical_2018}.

\begin{figure}[H]
\centering
\includegraphics[width=\textwidth]{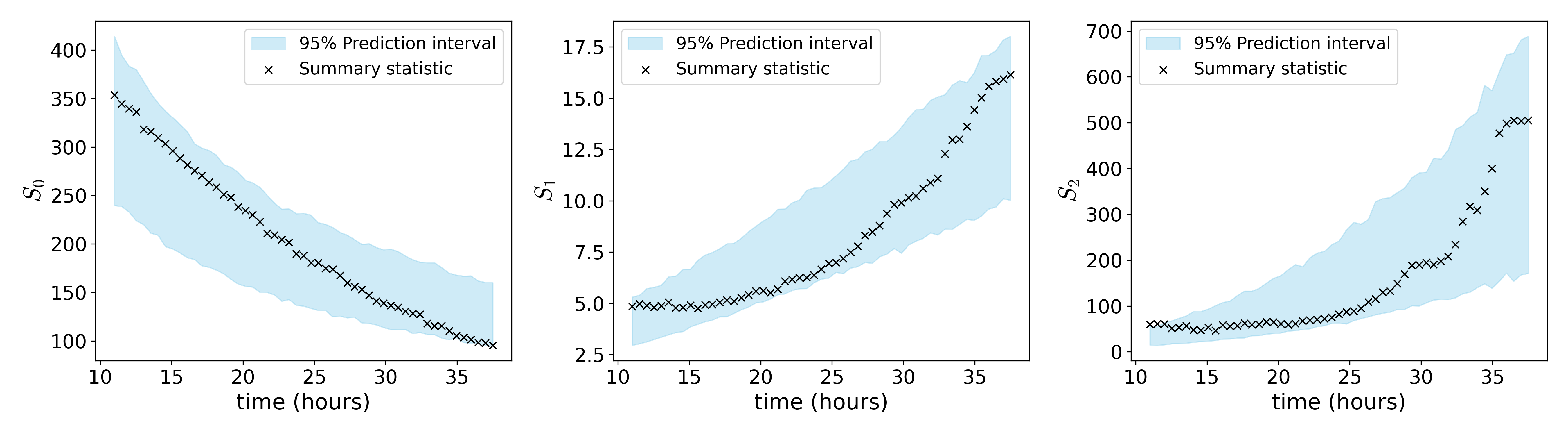}
\caption{Posterior predictive checks for the three summary statistics of the cluster size distribution when model variant $\mathcal{M}_{3,1}$ is fitted to the invasive cell data.
Each subplot shows the summary statistic values \rev{computed from experimental data} (black crosses) over time alongside the $95\%$ prediction interval (blue shaded region) generated from $1000$ posterior samples.}
\label{fig:INV_posterior_predictive}
\end{figure}

For the posterior predictive check, we simulated the full generative model $1000$ times using samples from the posterior distributions. For each posterior sample, we first simulated the noise-free model and generated the summary statistic outputs over the observation times. We then applied the observation noise model (Equation~(\ref{eqn:NoiseModel})) by multiplying each summary statistic value by a lognormal random factor with mean $0$ and variance $\sigma_k^2$, where $\sigma_k$ was drawn from the posterior for the noise parameter for the $k^{\text{th}}$ summary statistic. This procedure produces replicated datasets that incorporate both parameter uncertainty and measurement variability, enabling computation of $95\%$ prediction intervals by taking the empirical $2.5^{\text{th}}$ and $97.5^{\text{th}}$ percentiles of the simulated values. These intervals represent the range of values that the model predicts for the observed data under the posterior distribution and are visualised as shaded blue regions in Figure \ref{fig:INV_posterior_predictive}.
We note that the observed data falls within the posterior predictive intervals $98.1\%$ of the time when model variant $\mathcal{M}_{3,1}$ is fitted to the invasive cell data.

\begin{figure}[H]
\centering
\includegraphics[width=\textwidth]{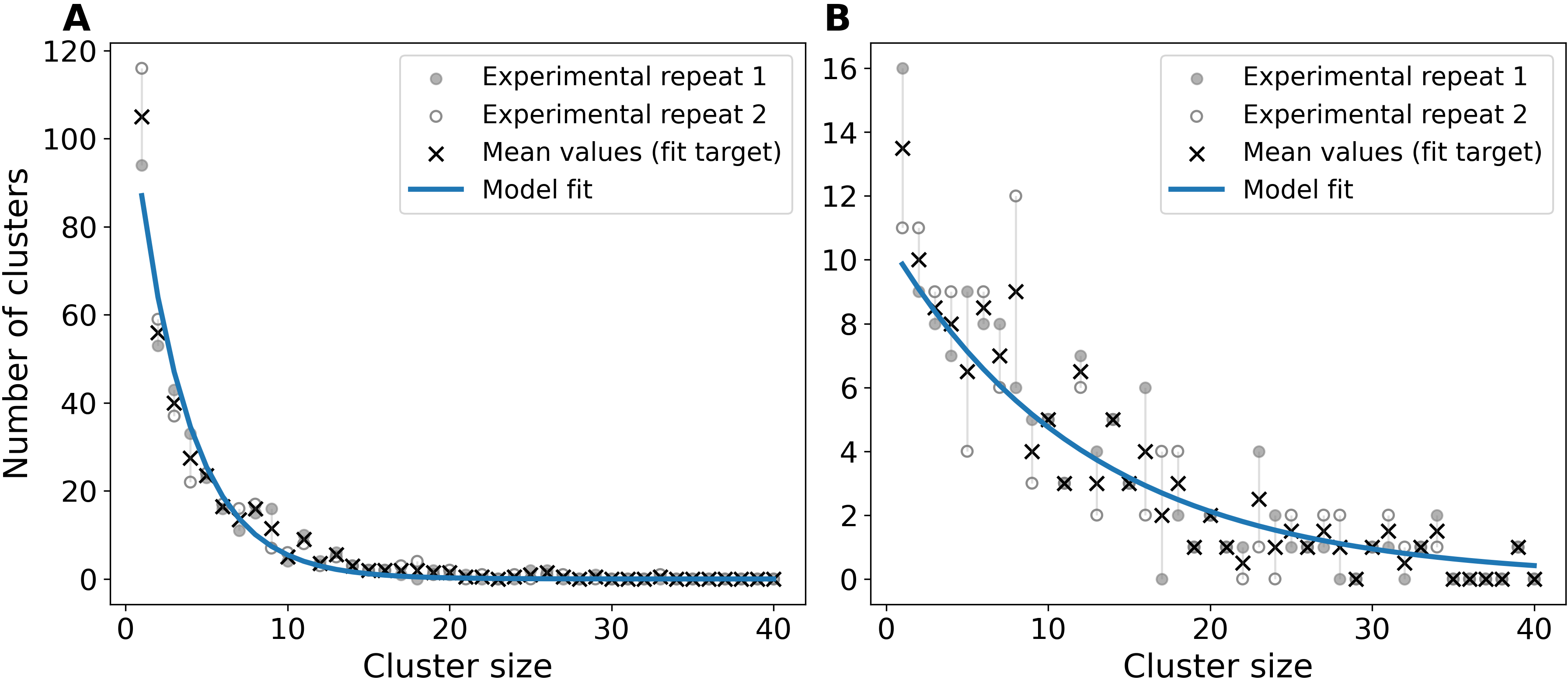}
\caption{Representative plot at (A) $t=11$ hours, and (B) $t=36$ hours of the distribution of cluster sizes in the experimental repeats for invasive cells and the distribution obtained from the maximum likelihood estimator (MLE) values for the parameters $\boldsymbol{\theta}_M = \{b = \rev{3.88\times10^{-4}} \,\text{clusters}^{-1}\,\rev{\text{h}^{-1}},\,
p = \rev{1.09\times 10^{-2}} \,\rev{\text{h}^{-1}},\,
q = 0 \,\rev{\text{h}^{-1}},\,
N_{\rev{0}} = 1101\}$ for model variant $\mathcal{M}_{3,1}$ when fitted to the summary statistics of the distribution. \rev{Black crosses indicate the mean values from the experimental data. Individual experimental repeats are also shown to highlight the variability in the observed data.}} 
\label{fig:INV_snapshot_fit}
\end{figure}

We also evaluate performance by simulation using the maximum likelihood estimator (MLE) values for the parameters $\boldsymbol{\theta}_M = \{b = \rev{3.88\times10^{-4}} \,\text{clusters}^{-1}\,\rev{\text{h}^{-1}},\,
p = \rev{1.09\times 10^{-2}} \,\rev{\text{h}^{-1}},\,
q = 0 \,\rev{\text{h}^{-1}},\,
N_{\rev{0}} = 1101\}$. The resultant cluster distributions are presented in Figure \ref{fig:INV_snapshot_fit} (blue line) together with the experimental data (black crosses) averaged across the experimental repeats. The plots show that the model accurately predicts the cluster size distribution at two distinct stages of the experiment. \rev{To quantify the agreement, we compute the root-mean-square error (RMSE) and the Pearson correlation coefficient, $r$, between the model predictions and experimental cluster size distributions at the representative time points shown in Figure~3. The RMSE values ($3.86$ at $t=11$ hours and $1.17$ at $t=36$ hours), correspond to only a small fraction of the observed cluster counts, while the corresponding correlation coefficients ($r=0.98$ and $r=0.95$) indicate that the model accurately captures the overall shape of the distributions. Together, these metrics indicate strong agreement between model predictions and experimental data.}
Further, the high level of coverage in the posterior predictive checks and the comparison of the model prediction to the cluster size distribution indicate that model variant $\mathcal{M}_{3,1}$ is a good fit to the data for the invasive cells.

\subsubsection*{Comparison of phenotypes}

To investigate how the cell phenotype influences clustering dynamics, we compare the clustering behaviours of proliferative and invasive cells in monoculture by fitting each of the model variants $\mathcal{M}_{\alpha,\beta}$ independently to each phenotype.
We present the parameter values associated with model variant $\mathcal{M}_{3,1}$ \rev{and the corresponding $95\%$ confidence intervals} in Table \ref{table:MLE-values-COMBO-M3,1}\del{ with the $95\%$ confidence intervals included in Supplementary Information Table S1}.
A notable difference between the proliferative and invasive cells relates to the coagulation rate, $b$. The MLE coagulation rates are $b =\rev{1.46 \times 10^{-4}} \,\text{clusters}^{-1}\,\rev{\text{h}^{-1}}$ for proliferative cells but $b =\rev{3.88 \times 10^{-4}} \,\text{clusters}^{-1}\,\rev{\text{h}^{-1}}$ for invasive cells. These results are consistent with experimental observations showing that invasive cells are more motile and adhesive than proliferative cells \citep{campbell_cooperation_2021}, and rapidly form larger clusters. 
The MLE value of the proliferation rate is slightly higher value for the proliferative cells $(p = \rev{1.43 \times 10^{-2}} \,\rev{\text{h}^{-1}})$ than for the invasive cells $(p = \rev{1.09 \times 10^{-2}} \,\rev{\text{h}^{-1}})$. These findings are consistent with experimental data showing that the doubling time of proliferative cells is shorter than that of invasive cells. \citep{campbell_cooperation_2021}.
\rev{The 95\% confidence intervals for $b$ and $p$ (Table~\ref{table:MLE-values-COMBO-M3,1}) indicate differences in the inferred parameter values between proliferative and invasive phenotypes. Although the posterior distributions exhibit some overlap, their centres are shifted relative to one another, suggesting systematic differences between the phenotypes.}

An identical procedure was used to fit all model variants to the proliferative and invasive cell datasets. We observed notable differences in the behaviour of the marginal posterior distributions for the shedding parameter $q$. When fit to the invasive cell data, every model variant that includes shedding produced a posterior distribution for $q$ that is sharply peaked at zero, indicating that including shedding into the model is not required to achieve a good fit to the data. A similar pattern occurs for some model variants when fitted to the proliferative cell data; however, for four model variants, $\mathcal{M}_{2,1},\, \mathcal{M}_{3,1},\, \mathcal{M}_{3,2},\, \mathcal{M}_{3,3}$, the posterior for $q$ is unimodal and centred away from zero, suggesting that the inclusion of shedding into the model improves the fit in these cases. These differences motivate a formal comparison of model variants using the AIC \citep{akaike_information_1998}.

\begin{figure}[H]
\centering
\includegraphics[width=\textwidth]{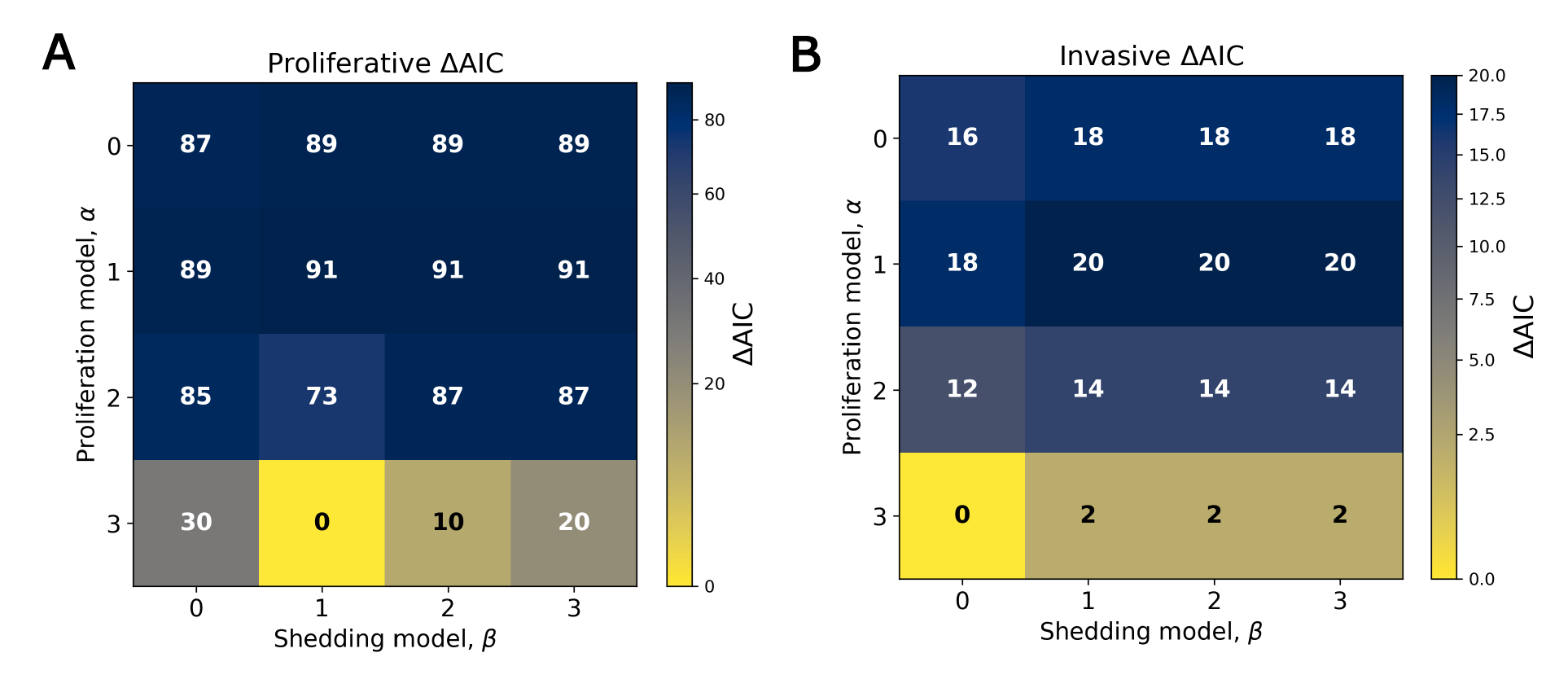}
\caption{$\Delta AIC$ values comparing different combinations of proliferation and shedding model variants for (A) proliferative and (B) invasive phenotype populations, assuming constant coagulation kernels. Each row corresponds to a different functional form for proliferation ($\alpha = 0$ for no proliferation, $\alpha = 1$ for constant proliferation, $\alpha = 2$ for surface area proliferation, and $\alpha = 3$ for cell number proliferation), while each column indicates the functional form for shedding $(\beta = 0,1,2,3)$. Lower $\Delta AIC$ values indicate better model performance.}
\label{fig:PRO_INV_AIC_values}
\end{figure}

The results in Figure \ref{fig:PRO_INV_AIC_values}A suggest that the model variant which best fits the proliferative cell data is $\mathcal{M}_{3,1}$. Similarly, Figure \ref{fig:PRO_INV_AIC_values}B suggests that the model variant which best fits the invasive cell data is $\mathcal{M}_{3,0}$. Thus, in both cases the best performing model variant has a proliferation rate proportional to the number of cells in a cluster, suggesting that for small clusters competition for space and nutrients does not limit proliferation.

Figure \ref{fig:PRO_INV_AIC_values} also shows that\rev{, for proliferative cells, model variants in which the posterior for the shedding parameter $q$ is concentrated away from zero exhibit lower $\Delta$AIC values, indicating that shedding improves the model fit in these cases.}
\del{ model variants in which the posterior for the shedding parameter $q$ is concentrated away from zero exhibit lower $\Delta AIC$ values. This pattern suggests that incorporating shedding provides a better fit to the data in these cases.}
Figure \ref{fig:PRO_INV_AIC_values} also shows that model variants in which the posterior for the shedding parameter $q$ is concentrated away from zero exhibit lower $\Delta AIC$ values. This pattern suggests that incorporating shedding provides a better fit to the data in these cases. However, this improvement only arises when shedding is paired with a more effective proliferation mechanism, specifically those in which proliferation scales with cell number or surface area, because the choice of proliferation mechanism has a larger overall influence on the $\Delta AIC$ value. This indicates that shedding plays a supporting role in proliferative cell cluster dynamics rather than being a primary driver; coagulation and proliferation are the dominant mechanisms. By contrast, for invasive cells, the posterior distribution for $q$ is sharply peaked at zero across all model variants, indicating that shedding does not contribute significantly to their clustering behaviour.

\rev{Coagulation in the model captures the effective rate at which clusters encounter and adhere to one another, while shedding represents cell detachment from established clusters. The combination of high coagulation rates and negligible shedding observed for invasive cells is consistent with the findings of \citealt{campbell_cooperation_2021}. Their study showed that the invasive phenotype forms clusters more readily than the proliferative phenotype and that, in human melanoma samples, the extent of clustering is positively associated with invasiveness (Figure 2E; \citealt{campbell_cooperation_2021}). This behaviour has been linked to increased expression of cell--cell adhesion molecules in invasive cells, which promotes both cluster formation and cluster stability. This differs from the classical epithelial--mesenchymal transition (EMT) paradigm often invoked in epithelial cancers, where invasive cells are expected to dissociate and migrate individually \citep{kalluri_basics_2009}. Melanoma does not generally follow a classical EMT programme and instead exhibits behaviours that are more consistent with collective migration. Within this framework, clusters can act as efficient invasive units, making the inferred combination of frequent coalescence and negligible shedding biologically plausible.}

\subsection{Uncovering mechanistic insights from heterogeneous clusters \label{section:ResultsCoculture}}

\begin{figure}[H]
\centering
\includegraphics[width=0.5\textwidth]{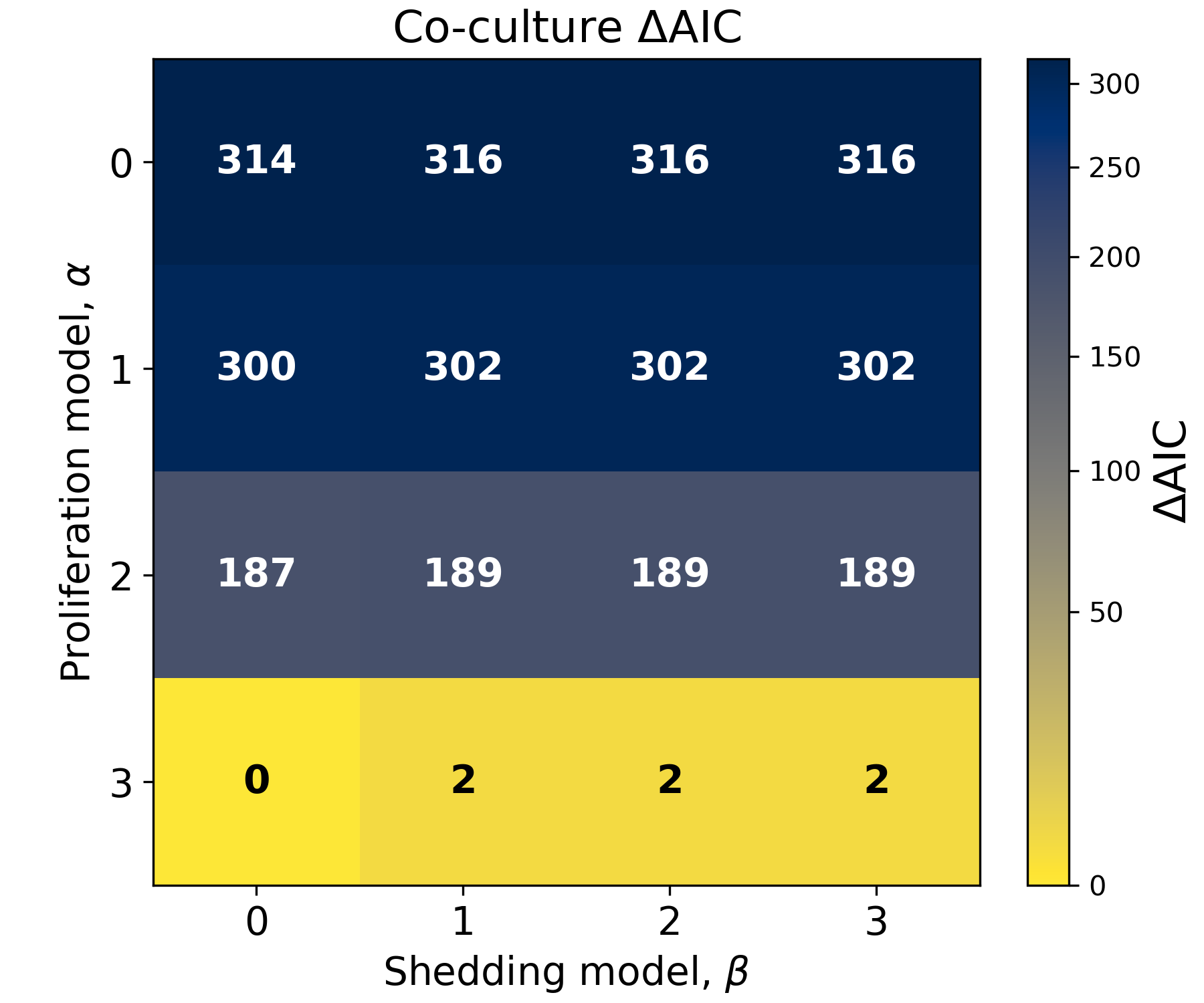}
\caption{$\Delta AIC$ values comparing different combinations of proliferation and shedding model variants for co-culture populations, assuming constant coagulation kernels. Each row corresponds to a different functional form for proliferation ($\alpha = 0$ for no proliferation, $\alpha = 1$ for constant proliferation, $\alpha = 2$ for surface area proliferation, and $\alpha = 3$ for cell number proliferation), while each column indicates the functional form for shedding $(\beta = 0,1,2,3)$. Lower $\Delta AIC$ values indicate better model performance.}
\label{fig:COMBO_AIC_values}
\end{figure}

We now analyse the data generated from the co-culture of proliferative and invasive phenotypes in order to understand the effect of phenotypic heterogeneity on clustering dynamics.
The model presented in Section \ref{section:HomogeneousModel} can be extended to model the co-culture experiments in a natural way, by considering the size and phenotypic composition of each cluster (that is, the number of cells of each phenotype within a given cluster). Details of this model are included in Supplementary Information Section S2. However, this model is not globally structurally identifiable without information on the phenotypic composition of each cluster, which cannot be extracted from the experimental data which is restricted to maximum intensity projections.
Therefore, we proceed by exploring the extent to which the model presented in Section \ref{section:HomogeneousModel}, which does not resolve phenotype, can provide mechanistic insight. Specifically, we fit the $16$ model variants $\mathcal{M}_{\alpha,\beta}$ to the summary statistics of the cluster size distribution data from the co-culture experiment. We performed model selection on these variants, and the detailed results can be found \rev{in Figure \ref{fig:COMBO_AIC_values}.} 

\begin{figure}[H]
\centering
\includegraphics[width=\textwidth]{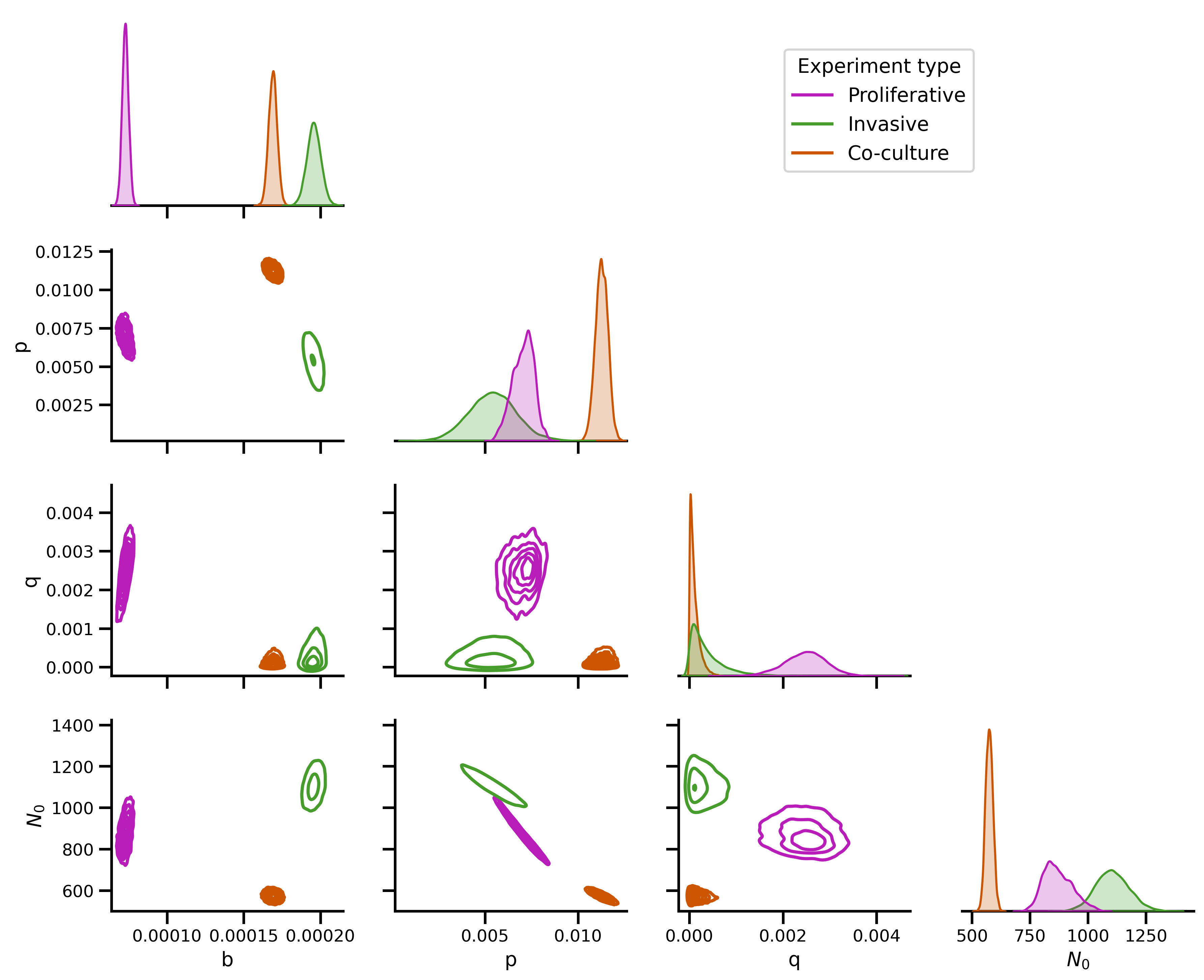}
\caption{\rev{Marginal and bivariate posterior distributions for model parameters $\boldsymbol{\theta_M} = \{b,p,q,N_{\rev{0}}\}$ when model variant $\mathcal{M}_{3,1}$ is fitted to data for each of the three experimental types. Plots are coloured by the experiment type (proliferative: magenta, invasive: green, and co-culture: orange).}}
\label{fig:PRO_INV_COMBO_Cell_num_cst_pairwise}
\end{figure}

The AIC \rev{values shown in Figure \ref{fig:COMBO_AIC_values} show that} model variant $\mathcal{M}_{3,0}$, which has a proliferation rate proportional to the number of cells in a cluster and no shedding, is the best performing model variant for the co-cultured data, matching the behaviour of the invasive phenotype, with cell number-based model variants for the proliferation term significantly outperforming the alternatives. %This suggests that the proliferation rate scales with the number of cells in a cluster, implying that each cell contributes equally to the overall growth dynamics. 
As model \rev{$\mathcal{M}_{3,1}$ is the best performing model variant for proliferative cells, and the best fitting model including shedding for both invasive cells and co-cultured cells, we now compare the behaviour across phenotypes using the fitted parameter values for this model. In particular, Table~\ref{table:MLE-values-COMBO-M3,1} reports the maximum likelihood estimates and Figure \ref{fig:PRO_INV_COMBO_Cell_num_cst_pairwise} presents the corresponding posterior and pairwise joint distributions for the monocultured and co-cultured data.} %under model variant $\mathcal{M}_{3,1}$, which assumes proliferation scales with cell number and includes constant-rate shedding. This model variant is chosen as it is the best performing model variant which includes shedding for co-cultured data and provides the best fit for proliferative cells (Figure~\ref{fig:PRO_INV_AIC_values}A). 
The choice of a model variant which includes shedding allows us to determine whether shedding contributes to clustering dynamics in each phenotype.

\begin{table}[]
\centering
\setlength{\tabcolsep}{5pt}
\renewcommand{\arraystretch}{1.2}
\begin{tabular}{|c|c|c|c|c|}
\hline
\textbf{Model} $\mathcal{M}_{3,1}$ & $b \,(\text{clusters}^{-1}\,\text{h}^{-1})$ & $p \,(\text{h}^{-1})$ & $q \,(\text{h}^{-1})$ & $N_{\rev{0}}$ \\
\hline

Proliferative cells 
& $\begin{array}{c} 1.46\times10^{-4} \\ {\footnotesize [1.45,1.47]\times10^{-4}} \end{array}$
& $\begin{array}{c} 1.43\times10^{-2} \\ {\footnotesize [1.42,1.44]\times10^{-2}} \end{array}$
& $\begin{array}{c} 4.56\times10^{-3} \\ {\footnotesize [4.31,4.82]\times10^{-3}} \end{array}$
& $\begin{array}{c} 847 \\ {\footnotesize [841,854]} \end{array}$
\\
\hline

Invasive cells 
& $\begin{array}{c} 3.88\times10^{-4} \\ {\footnotesize [3.78,3.98]\times10^{-4}} \end{array}$
& $\begin{array}{c} 1.09\times10^{-2} \\ {\footnotesize [9.72,11.98]\times10^{-3}} \end{array}$
& $\begin{array}{c} 0 \\ {\footnotesize [0,1.11]\times10^{-3}} \end{array}$
& $\begin{array}{c} 1101 \\ {\footnotesize [1065,1139]} \end{array}$
\\
\hline

Co-culture cells 
& $\begin{array}{c} 3.37\times10^{-4} \\ {\footnotesize [3.33,3.41]\times10^{-4}} \end{array}$
& $\begin{array}{c} 2.24\times10^{-2} \\ {\footnotesize [2.22,2.27]\times10^{-2}} \end{array}$
& $\begin{array}{c} 0 \\ {\footnotesize [0,3.18]\times10^{-4}} \end{array}$
& $\begin{array}{c} 575 \\ {\footnotesize [568,582]} \end{array}$
\\
\hline

\end{tabular}
\caption{\rev{Comparison of the MLE values (top value in a box) and 95\% confidence intervals for MLE values (bottom values in a box) for model parameters $\boldsymbol{\theta}_M = \{b,p,q,N_{\rev{0}}\}$ across cell lines for model variant $\mathcal{M}_{3,1}$ when fitted to the monocultured proliferative and invasive datasets as well as the co-cultured data.}}
\label{table:MLE-values-COMBO-M3,1}
\end{table}

The posterior distributions and MLE values provide insight into the behaviour of heterogeneous cell clusters. As illustrated in Figure \ref{fig:PRO_INV_COMBO_Cell_num_cst_pairwise}, the negative correlation between parameters $p$ and $N_{\rev{0}}$ persists across all conditions. Table \ref{table:MLE-values-COMBO-M3,1} shows that the MLE value for the coagulation rate of the co-cultured data lies between the values for the proliferative and invasive monocultures. However, it is closer to that of the invasive cells, suggesting that coagulation dynamics within mixed-phenotype clusters are dominated by invasive cells. 
In contrast, the MLE value of the proliferation rate for the co-cultured data is notably higher than the values for the either monoculture. This suggests that there may be a mutually beneficial effect of co-clustering which increases the net proliferation rate.
The inclusion of shedding in the current ODE framework does not significantly improve model performance for co-cultured cells, as evidenced by the posterior for the shedding parameter, which is sharply peaked at zero and rapidly decays. %(see Supplementary Information Figure S6A). 
This mirrors the behaviour of invasive cells and suggests that shedding is not essential to explain the co-cultured data within a spatially averaged model.

Together the results presented in this work\del{ show that it is possible to distinguish the three experimental phenotypic conditions from the data} \rev{suggest that the three experimental phenotypic conditions exhibit distinct clustering behaviours, and that the estimated model parameters and best-performing model formulations provide insight into the mechanisms driving cluster formation.} \del{and comparing functional forms that may explain the mechanisms driving cluster formation. We conclude that the model serves as a practical diagnostic tool for assessing cluster formation dynamics,}\rev{We conclude that the model provides a quantitative framework for assessing cluster formation dynamics,}
is well-aligned with the characteristics of the data across all three experimental conditions, and\del{ provides a baseline for detecting} \rev{may serve as a baseline for identifying} differences in clustering behaviour between phenotypes and between monoculture and co-culture experiments.

% Together the results displayed in this work show that it is possible to distinguish the three experimental phenotypic conditions from the data and to compare functional forms that may explain the mechanisms driving cluster formation. We conclude that the model serves as a practical diagnostic tool for assessing cluster formation dynamics, is well-aligned with the characteristics of the data across all three experimental conditions, and provides a baseline for detecting differences in clustering behaviour between phenotypes and between monoculture and co-culture experiments.

\section{Discussion \label{section:Discussion}}

This study presents a novel application of structured ODE models to investigate the clustering behaviour of melanoma cells, with a particular focus on the differences between melanoma cell phenotypes. By integrating \textit{in vitro} imaging data with mechanistic modelling, we developed a quantitative framework that captures the key biological processes of coagulation, proliferation, and shedding and reveals how the relative contributions of these processes differ between proliferative and invasive phenotypes. These two phenotypes represent ``bookends" of how cancer cells behave in the classic “go or grow” model (a cell is either proliferative or invasive but not both). Significant nuances to this “go or grow” model have emerged, with single-cell RNA-seq revealing that there are many discrete phenotype states in most cancers including melanoma \citep{belote_human_2021,lim_single-cell_2024, lumaquin-yin_lipid_2023,rambow_melanoma_2019, tirosh_cancer_2024}. In this study we have focussed on two extreme phenotypes (proliferative and invasive), given the detailed data available for these phenotypes from prior work \citep{campbell_cooperation_2021}.

A total of $16$ mechanistic model variants were compared to investigate the processes governing cell proliferation and shedding. When fitted to monoculture data, the best fitting model variant predicts a significantly higher coagulation rate for the invasive phenotype than compared to the proliferative phenotype, and a proliferation rate for the proliferative phenotype that is slightly higher than for the invasive phenotype.
These phenotypic differences align with known gene expression data for the two phenotypes. Genes in the TFAP2 family (namely TFAP2A and TFAP2E), play essential roles in neural crest and melanocyte fate specification during development \citep{de_croze_reiterative_2011,hoffman_tfap2_2007,li_redundant_2007,luo_transcription_2002,seberg_tfap2_2017,van_otterloo_differentiation_2010}. They have been shown to be a master regulator of phenotypic heterogeneity in melanoma \citep{campbell_cooperation_2021}, as they are upregulated in invasive cells and downregulated in proliferative cells \citep{hoek_metastatic_2006,rambow_new_2015,tirosh_dissecting_2016,verfaillie_decoding_2015}. This leads to increased cadherin gene expression and cell adhesion in invasive cells, whilst also upregulating CDK genes in proliferative cells, promoting faster cell cycle progression. 
The consistency between model predictions and\del{ biological data—especially gene expression patterns—demonstrates the model’s capacity to accurately reflect phenotype-specific dynamics.} \rev{known biological data (such as gene expression patterns) supports the model's ability to capture phenotype-specific dynamics.}
% The consistency between model predictions and biological data—especially gene expression patterns—demonstrates the model’s capacity to accurately reflect phenotype-specific dynamics. 

Among the proliferation models tested, the cell number-dependent model variants (denoted $\mathcal{M}_{3,\beta}$ for shedding model $\beta$) consistently provided the best fit, suggesting that proliferation is proportional to the total number of cells within a cluster. This suggests that the clusters are still sufficiently small that spatial constraints and nutrient limitations are negligible. In such contexts, cell–cell interactions and collective signalling are likely to play a prominent role. Although cluster shedding is observable in the data, it has less impact on model performance. Shedding only improves the fit in a subset of proliferative cell model variants and, even then, only when paired with better performing proliferation mechanisms, highlighting its role as a secondary contributor compared to coagulation and proliferation.

%%%%%%%%%%%%

The natural extension of the homogeneous model to a heterogeneous model is not globally structurally identifiable with the available data. Therefore, we analysed the clustering behaviour observed in co-culture experiments as for homogeneous clusters.  
We observe a hybrid coagulation behaviour in co-culture, with coagulation rates falling between those for the monocultures but closer to those of the invasive phenotype, indicating that invasive cells dominate the coagulation dynamics within mixed phenotype experiments. 
The proliferation rate observed in co-culture experiments exceeds that of both proliferative and invasive monoculture experiments, suggesting a mutually beneficial interaction between the two phenotypes that enhances overall proliferative activity. A possible explanation for this behaviour is the secretion of growth-promoting factors into the shared microenvironment. In co-culture systems, cells often release distinct sets of signalling molecules that can influence the behaviour of neighbouring cells \citep{barkley_cancer_2022, pozniak_tcf4-dependent_2024}. Although the specific factors secreted by proliferative and invasive phenotypes remain unidentified, it is plausible that their combined secretome creates an environment that is more favourable for proliferation than either phenotype alone. A similar phenomenon has been observed in other systems—for example, keratinocytes have been shown to enhance melanoma cell proliferation via secretion of EDN3 \citep{lahav_endothelin_1996, saldana-caboverde_roles_2010}. 
Another possible explanation for the increased proliferation rate in co-culture experiments is the creation of physical niches \citep{nowosad_perivascular_2023} by invasive phenotype cells that allow proliferative phenotype cells to proliferate more readily. 
The increased proliferative behaviour of the co-cultures underscores the importance of considering cell–cell interactions when interpreting proliferation dynamics in heterogeneous systems.
These results also demonstrate that homogeneous models, when carefully interpreted, can serve as effective diagnostic tools for detecting emergent behaviours in heterogeneous systems. While this model does not explicitly track phenotypic composition, it enables us to extract mechanistic insights from the available co-culture experiments. 

Although the model provides useful insight into clustering dynamics, it does not capture the spatial organisation within clusters or interactions between different cell phenotypes. For example, in heterogeneous clusters invasive phenotype cells tend to occupy the centre of each cluster, whereas proliferative phenotype cells form a partial outer layer. These limitations highlight the importance of model‑guided experimental design: integrating mathematical modelling with experiment can help direct future studies towards collecting additional information, such as the phenotypic composition within clusters, that would support the development of more detailed and biologically informed models.
In addition, integrating the imaging data with the available RNA-sequencing (both bulk and single-cell) data would allow for far greater nuance on how phenotype is linked to gene expression.  

In summary, by combining mathematical modelling with \textit{in vitro} data, we have shown that melanoma cluster formation is driven by coagulation and cell number-dependent proliferation, with distinct behaviours between proliferative and invasive phenotypes. Although the model does not explicitly track phenotypic composition within clusters,\del{ it can still distinguish differences in clustering behaviour} \rev{it can still reveal differences in clustering behaviour} between monocultures and co-cultures. This ability provides a practical tool for detecting emergent dynamics in co-cultured systems and offers a foundation for understanding early metastatic processes.
This work also highlights and motivates the importance of integrating richer experimental data and more detailed modelling frameworks to fully capture the complexity of co-cultured systems.

\backmatter

\bibliography{paper-bibliography}% common bib file

\section*{Supplementary information}

Supplementary Information is available online and provides additional methodological details and supporting analyses.

\section*{Acknowledgements}

Funding: N.J.S. is supported by funding from the Engineering and Physical Sciences Research Council (EPSRC) [grant number EP/S024093/1], N.J.S, H.M.B, and R.M.W are all supported by the Ludwig Institute for Cancer Research (Oxford Branch),  R.E.B. is supported by a grant from the Simons Foundation (MP-SIP-00001828).

\section*{CRediT authorship contribution statement}
N.J.S.: Writing – original draft, Formal analysis, Software, Methodology, Conceptualisation, Investigation, Data curation, Visualisation.
H.M.B.: Writing – review and editing, Supervision, Methodology, Validation, Conceptualisation.
R.E.B.: Writing – review and editing, Supervision, Methodology, Validation, Conceptualisation.
R.M.W.: Resources, Investigation, Data curation, Validation, Writing – review and editing, Supervision, Conceptualisation.

\section*{Ethics declarations}

\noindent Competing Interests: The authors declare that they have no competing interests.

\vspace{0.5em}

\noindent Data Availability: The experimental imaging data analysed in this study are available from the original publication \citep{campbell_cooperation_2021}. Processed cluster size distributions and summary statistics generated for model fitting is available at https://github.com/njs59/ODE-phenotype-clustering.

\vspace{0.5em}

\noindent Code Availability: All code used for model simulation, Bayesian inference, and analysis are available at https://github.com/njs59/ODE-phenotype-clustering.

\end{document}